\documentclass[a4paper,12pt]{article}
\usepackage[utf8]{inputenc}
\usepackage{amsmath}
\usepackage{amssymb}
\usepackage{amsfonts}
\allowdisplaybreaks
\numberwithin{equation}{section}

\def\Im{\operatorname{Im}}

\begin{document}

\title{Integrable equations with negative evolution numbers.}
\author{A.~K.~Pogrebkov\\
Steklov Mathematical Institute of\\ Russian Academy of Sciences, Moscow, Russiaof times; hierarchies; dimensional reductions.} 

\maketitle

\begin{abstract}
Two integrable systems are constructed in a $(2+1)$-dimensional space. Each of these systems involves two evolutions with negative numbers.
\end{abstract}
 
\section{Introduction}
It is well known that the times of hierarchies $(2+1)$-dimensional integrable systems can be conveniently introduced by means of commutators,  see \cite{akp2008a}. Say in the case of the Davey--Stewartson (DS) hierarchy, \cite{DS}, we have two sets ot times:
\begin{equation}
B_{t_m}=[A^m,B],\qquad B_{x_{m}}=i[\sigma A^{m},B], \label{tmds} 
\end{equation}
where
\begin{equation}
 \sigma^2=1,\qquad [\sigma,A]=0,\qquad \sigma B+B\sigma=0.
 \label{sigma}
\end{equation}
Here $m\in\mathbb{Z}$, $m\geq1$, $A$ and $B$ are arbitrary elements of an associative algebra $\mathcal{A}$ with unity $I$. We can consider elements $A$, $B$, and $\sigma$ as $2\times2$ matrices, $A$ to be proportional to the unity matrix $I$, $B$ to be the off-diagonal matrix and $\sigma=\sigma_3$, the Pauli matrix. These relations are not mutually independent. For instance, for any $A$ and $B$ we have identities
\begin{equation}
[A^{2m},B]=\sigma[A^m,[\sigma A^{m},B]],\label{comm0} 
\end{equation}
see \cite{akp2020b}. Commutator identities of this type lead due to (\ref{tmds}) to linear differential equations for $B$. Thus for $m=1$ we get
\begin{equation}
B_{t_{2}}+i\sigma B_{t_1x_1}=0,\label{DS} 
\end{equation}
i.e., a version of the linearized DS equation (the lowest one in the hierarchy). Similar relations exist for the higher values of $m$ in (\ref{tmds}), see \cite{akp2020b}.

In order to raise a linear differential equation to a nonlinear integrable one, we must select one of the variables $t_m$, $x_m$ as the \textsl{leading}, in a sense that all other derivatives will be given as differential operators with respect to this one. If we choose $t_1$ as leading then by means of (\ref{tmds}) for $m=1$ we set 
\begin{equation}
AB=(\partial_{t_1}+z)B,\qquad BA=zB,\label{AB}  
\end{equation}
where we took into account that operator $A$ is defined up to an arbitrary parameter $z\in\mathbb{Z}$. Thus by (\ref{tmds}) and (\ref{AB}) we obtain
\begin{equation}
B_{t_m}=[(z+\partial_{t_1})^{m}-z^m]B,\qquad
B_{x_m}=i\sigma[(z+\partial_{t_1})^{m}+z^m]B.\label{Btm}
\end{equation}
So we see that all evolutions, with the exception of the leading one, are expressed through the latter. Moreover, the effect of all these evolutions depends on the parameter $z$, introduced as an ambiguity in the construction of $A$ by  (\ref{tmds}) for $m=1$. As shown in \cite{akp2016}, it is this complex parameter that makes it possible to construct the Inverse Problem.

In \cite{akp2008a}, \cite{akp2016} we have considered the inverse problem based on the dressing procedure, where elements $A$ and $B$ of the algebra 
$\mathcal{A}$ are implemented as (pseudo-)differential operators with symbols depending on $t_1$, several other times from (\ref{tmds}) and $z$. The symbol for the composition of two operators is defined as
\begin{equation}
(FG)(t,x,z)=\dfrac{1}{2\pi}\int dp\int dy\,F(t,x,z+ip)e^{ip(t_1-y_1)}G(y,x,z),\label{comp0}
\end{equation}
if the integral is well defined in the sense of distributions. Here we denote $t=\{t_1,t_2,\ldots\}$, $x=\{x_1,x_2,\ldots\}$, $y=\{y_1,t_2,t_3,\ldots\}$ and $\ldots$ denote some finite subsets of times. Symbol of the unity operator $I$ is the unity $2\times2$ matrix  and symbol of operator $A$ is 
\begin{equation}
A=z,\label{Az0}
\end{equation}
where we omitted factor $I$. We see that time $t_1$ plays the special role, indeed. Due to (\ref{comp0}) we have for any pseudo-differential operator $F$ equalities:
\begin{equation}
(A^mF)(t,x,z)=(z+\partial_{t_1})^{m}F(t,x,z),\quad(FA^m)(t,x,z)=z^{m}F(t,x,z),\label{AF0}
\end{equation}
where $A^m$ denotes the $m$-th power in the sense of composition (\ref{comp0}). In particular we have for $m=1$: 
$(AF)(t,x,z)=F_{t_1}(t,x,z)+zF(t,x,z)$, $ (FA)(t,x,z)=zF(t,x,z)$, so that
\begin{equation}
F_{t_1}=[A,F].\label{AF1}                
\end{equation}
Thus due to (\ref{comp0}) we have this equality not for the scattering data $B$ for $m=1$ only, but for any $F$. Moreover, variable $z$ cancels out from action of the derivative with respect to $t_1$, as it must be for the leading evolution. Notice that (\ref{AF1}) can be understood as a kind of ``normal ordering'',
\begin{equation}
AF=F_{t_1}+FA,\label{norm1} 
\end{equation}
that enables to shift all operators $A^{m}$  to the rightmost position.

On this set of operators we introduce 
$\overline\partial$-derivative with respect to the complex variable $z$, $F\to\bar\partial{F}$. In terms of symbols, it is defined, see \cite{akp2008a}, as
\begin{equation}
(\overline{\partial}F)(t,x,z)=\dfrac{\partial {F(t,x,z)}}{\partial\overline{z}},\label{dbar3}
\end{equation}
where the derivative is understood in the sense of distributions. By virtue of (\ref{Az0}), we get the following equality: 
\begin{equation}
\overline{\partial}A=0,\label{A0}                    
\end{equation}
which plays a crucial role in the subsequent construction.
  
The dressing operator $K$ with the symbol 
$K(t,x,z)$, that also depends on some times of the sets $\{t_m\}$ and $\{x_m\}$, is defined, \cite{akp2008a}, by means of the $\overline\partial$-problem
\begin{equation}
\overline\partial K=KB,\label{dbar_1}
\end{equation}
where the composition on the right side is understood in the sense of (\ref{comp0}). The solution $K$ of the equation (\ref{dbar_1}) is normalized by the asymptotic condition on the symbol
\begin{equation}
K(t,x,z)\to I,\quad z\to\infty,\label{dbar_2} 
\end{equation}
where $I$ denotes $2\times2$ unity matrix. We assume unique solvability of the $\overline\partial$-problem (\ref{dbar_1}), (\ref{dbar_2}).  This condition is not burdensome, since we are building Lax pairs, and the effectiveness of these pairs can be tested directly. 

If we associate unity dimensionality to $z$, then all derivatives $\partial_{t_m}$, $\partial_{x_m}$ in (\ref{tmds}) get dimensionality $m$. In presentation above value of $m$ was positive, so we refer to such $t_m$ and $x_m$ as times with positive numbers. This is the standard case for the Inverse Problem because of (\ref{Az0}) and majority of Integrable hierarchies involves times with positive numbers. In fact the only example of the system with negative time gives the Toda chain, see \cite{man1976,mikh}. In \cite{akp2021a}, \cite{akp2021b} we constructed new integrable equations with one time with negative number. Here we present construction of the integrable system with two evolutions with negative numbers.

\section{System with two evolutions with negative numbers.}
\subsection{Commutator identities and linear equations.}

Let us consider two versions of 2-dimensional Toda chain defined by the commutator identities
\begin{align}
&[A,[A^{-1},B]]+ABA^{-1}-2B+A^{-1}BA=0,\label{b8}\\
&[A,[\sigma A^{-1},B]]-\sigma ABA^{-1}+\sigma A^{-1}BA=0.\label{b80}
\end{align}
Thus we extend equalities (\ref{tmds}) to $m=-1$, i.e., we set
\begin{equation}
B_{t_{-1}}=[A^{-1},B],\qquad B_{x_{-1}}=i[\sigma A^{-1},B],\label{b3} 
\end{equation}
where $\sigma$ is defined in (\ref{sigma}). Moreover,
these identities suggest introduction of the discrete ``time'' $n\in\mathbb{Z}$ in $B$. So that now symbol of the pseudo-differential operator $F\in\mathcal{A}$ depends on variables $t_1$, $t_{-1}$, $x_{-1}$, $n$, and $z$. We also denote shift of the discrete variable $n$ as
\begin{equation}
F^{(1)}(t_1,t_{-1},x_{-1},n,z)=
F(t_1,t_{-1},x_{-1},n+1,z).\label{b7} 
\end{equation}
We see (cf.\ \cite{akp2008b}) that evolution with the negative number leads to introduction of discrete variable. Evolutions of such variables are given as commutators also, but in the  group sense, i.e., by similarity transformations. We introduce evolution of the Scattering Data with respect to the discrete variable by means of equalities
\begin{equation}
B^{(1)}=ABA^{-1},\qquad B^{(-1)}=A^{-1}BA, \label{b6}
\end{equation}
so that (\ref{b8}), (\ref{b80}) prove that $B$ obeys linear differential-difference equations
\begin{align}
&B_{t_1t_{-1}}+B^{(1)}-2B+B^{(-1)}=0,\label{b10}\\
&B_{t_1x_{-1}}-i\sigma B+i\sigma B^{(-1)}=0,\label{b100}
\end{align}
where we use (\ref{AF1}).

Eqs.\ (\ref{b10}) and (\ref{b100}) are linearized versions of the Toda chain. Every of them involves only one time with negative number. Thus we have to exclude either commutator $[A,B]$, or shift $ABA^{-1}$. So we have a new commutator identity
\begin{equation}
 [A,[\sigma A^{-1},[\sigma A^{-1},B]]-
 [A^{-1},[A^{-1},B]]]+4[A^{-1},B]=0, \label{b1}
\end{equation}
that due to (\ref{AF1}) and (\ref{b3}) gives the linear differential equation
\begin{equation}
\bigl(B_{x_{-1}x_{-1}}+B_{t_{-1}t_{-1}}\bigr)_{t_1}-4B_{t_{-1}}=0,
\label{b5}
\end{equation}
with independent variables $t_1$, $t_{-1}$, and $x_{-1}$, where the last two have negative numbers.

On the other hand we can exclude commutator $[A,B]$ by means of the similarity transformation (\ref{b6}). In this way we arrive to the commutator identity
\begin{equation}
[A^{-1},ABA^{-1}]+\sigma[\sigma A^{-1},ABA^{-1}]+[A^{-1},B]-\sigma[\sigma A^{-1},B]=0,\label{b101}
\end{equation}
and definitions (\ref{b3}) and (\ref{b6}) show that operator $B$ obeys linear dif\-fe\-ren\-ti\-al-difference equation
\begin{equation}
B^{(1)}_{t_{-1}}-i\sigma B^{(1)}_{x_{-1}}+B^{}_{t_{-1}}+i\sigma B^{}_{x_{-1}}=0.\label{b102}
\end{equation}
Thus we have one discrete independent variable $n$ and two evolutions $t_{-1}$ and $x_{-1}$ with negative numbers. It is necessary to understand if equation (\ref{b102}) admit nonlinearization to the integrable one. For this purpose we have to consider a special version of pseudo-differential operators.

\subsection{Pseudo-differential operators.}
Following Introduction, such operators denoted as $F$, $G$, etc., are defined by the corresponding symbols $F(t_1,t_{-1},x_{-1},n,z)$, $G(t_1,t_{-1},x_{-1},n,z)$, etc., where $z\in\mathbb{C}$ is an arbitrary complex parameter. Thanks to (\ref{comp0}) variable $t_1$ is the leading time variable. Symbol of composition of two pseudo-differential operators is given by means of a pointwise product of their symbols with respect to variables $t_{-1}$, $x_{-1}$,  and $n$. The symbol of the unity operator $I$ is the unity $2\times2$ matrix, the symbol of the operator $A$ is given in (\ref{Az0}). In the following, we consider this space as space of tempered distributions in the arguments of their symbols.

Action of operations (\ref{AF0}) are more transparent in terms of the Fourier transform with respect to $t_1$:
\begin{equation}
\widetilde{F}(p,z)=\dfrac{1}{2\pi}\int dt_1\,e^{-ipt_1}F(t_1,z)\label{f1} 
\end{equation}
(we omit other variables for convenience), so that (\ref{comp0}) takes the form
\begin{equation}
\widetilde{(FG)}(p,z)=\int dp'\widetilde{F}(p-p',z+ip') \widetilde{G}(p',z),\label{comp1} 
\end{equation}
where we omit inessential variables.
Correspondingly, by (\ref{Az0}) and (\ref{AF0}) we have
\begin{align}
&\widetilde{A}(p,z)=z\delta(p),\label{f3}\\ 
&\widetilde{(A^{-1}F)}(p,t_{-1},x_{-1},n,z)=
\dfrac{\widetilde{F}(p,t_{-1},x_{-1},n,z)}{z+ip},\label{f11}\\
&\widetilde{(FA^{-1})}(t_1,t_{-1},x_{-1},n,z)=
\dfrac{\widetilde{F}(p,t_{-1},x_{-1},n,z)}{z}.\label{AF-1}
\end{align}

In terms of the Fourier transform (\ref{b1}) relations (\ref{b3}), (\ref{b6}), and (\ref{AF1}) sound as
\begin{align}
&\widetilde{B_{t_{-1}}}=\biggl(\dfrac{1}{z+ip}-\dfrac{1}{z}\biggr)\widetilde{B},\qquad
\widetilde{B_{x_{-1}}}=i\sigma\biggl(\dfrac{1}{z+ip}+\dfrac{1}{z}\biggr)\widetilde{B},\label{f23}\\
&\widetilde{B^{(1)}}=\dfrac{z+ip}{z}
\widetilde{B},\qquad \widetilde{B^{(-1)}}=\dfrac{z}{z+ip}\widetilde{B},\qquad
\widetilde{B_{t_1}}=ip\widetilde{B}.\label{f22}
\end{align}

Equations (\ref{Btm}) and (\ref{b6}) allow us to find a representation for the general form of the symbol of operator $B$:
\begin{align}
B&(t_1,t_{-1},x_{-1},n,z)=\int dp\,f(p,z)
\biggl(\dfrac{z+ip}{z}\biggr)^{n}\times\nonumber\\&\times\exp\Bigl(ipt_1+((z+ip)^{-1}-z^{-1})t_{-1}+
i\sigma((z+ip)^{-1}+z^{-1})x_{-1}\Bigr),\label{Bf} 
\end{align}
where $f(p,z)$ is an off-diagonal $2\times2$ matrix with elements that are arbitrary functions of their variables that guarantees the convergence of the integral and the boundedness of the limits when $t_1$ tends to infinity. There are two ways to achieve this. The first one is given by choice $f(p,z)=\delta(p+2z_{\Im})g(z)$. Then (\ref{Bf}) takes the form
\begin{align}
B(t_1,t_{-1},&x_{-1},n,z)=g(z)
\biggl(\dfrac{\overline{z}}{z}\biggr)^{n}\times\nonumber\\&\times\exp\biggl((\overline{z}-z)t_1+\biggl(\dfrac{1}{\overline{z}}-\dfrac{1}{z}\biggr)t_{-1}+
i\sigma\biggl(\dfrac{1}{\overline{z}}+\dfrac{1}{z}\biggr) x_{-1}\biggr),\label{Bf2} 
\end{align}
where $g(z)$ inherits properties of matrix $f(p,z)$ above. 
For another way to impose condition of boundedness on the integral in (\ref{Bf}) see \cite{akp2020b}. 

\subsection{Evolutions on the dressing operator.}
We define Dressing Procedure by means of the $\overline\partial$-problem (\ref{dbar_1}), (\ref{dbar_2}) with composition in the r.h.s.\ given in (\ref{comp0}). Thus we derive that evolutions (\ref{AF1}), (\ref{b3}), and (\ref{b6}) result in the following evolutions of the dressing operator:  
\begin{align}
&\overline\partial K_{t_1}=K_{t_1}B+K[A,B],\label{Kt}\\
&\overline\partial K_{t_{-1}}=K_{t_{-1}}B+K[A^{-1},B],
\label{Kt-1}\\
&\overline\partial K_{x_{-1}}=K_{x_{-1}}B+iK[\sigma A^{-1},B],
\label{Kx-1}\\
&\overline\partial K^{(1)}=K^{(1)}ABA^{-1}.\label{K}
\end{align}
In \cite{akp2020b} we proved that evolutions with positive numbers of times of the dressing operator mutually commute  under condition of unique solvability of the Inverse Problem. Here this commutativity also follows from commutativity of derivatives of the scattering data $B$ due to (\ref{AF1}), (\ref{b3}), and (\ref{b6}). But in order to ensure that the symbol of operator $B$ belongs to the space of tempered distributions with respect to $z$, we impose the condition that $g(0)=0$ in (\ref{Bf2}).

Notice that (\ref{Kt}) is nothing but (\ref{norm1}) and can be used as ``normal ordering'' in relations of the kind $A^{n}K=\sum_{m=0}^{n}C^{n}_{m}(\partial^{n-m}_{t_1}K)A^{m}$. Next, consider (\ref{Kt-1}). Multiplying by $A$ from the right we get $\overline\partial (K_{t_{-1}}A)=K_{t_{-1}}BA+KA^{-1}BA-KB$, where in the l.h.s.\ we used (\ref{A0}). Using in the last term (\ref{dbar_1}), we get
$\overline\partial (K_{t_{-1}}A+K)=(K_{t_{-1}}A+K)A^{-1}BA$, or thanks to (\ref{b6}) $\overline\partial (K_{t_{-1}}A+K)=(K_{t_{-1}}A+K)B^{(-1)}$. Now shift with respect to the discrete variable $n$ gives $\overline\partial (K^{(1)}_{t_{-1}}A+K^{(1)})=(K^{(1)}_{t_{-1}}A+K^{(1)})B$, i.e., combination $K^{(1)}_{t_{-1}}A+K^{(1)} $ obeys the original $\overline\partial$-equation (\ref{dbar_1}), while the asymptotic condition is different to (\ref{dbar_2}). We assume that besides (\ref{dbar_1}), the symbol of the operator $K$ admits an asymptotic expansion 
\begin{align}
K(t_1,t_{-1},x_{-1},n,z)&=I+\dfrac{u(t_1,t_{-1},x_{-1},n)}{z}+\nonumber\\
&+\dfrac{v(t_1,t_{-1},x_{-1},n)}{z^2}+o(z^{-2}),\qquad z\to\infty,\label{dbar4} 
\end{align}
where $u(t_1,t_{-1},x_{-1},n)$ and $v(t_1,t_{-1},x_{-1},n)$ are $2\times2$ matrices. Their elements are independent of $z$, so they are multiplication operators in the sense of the composition law (\ref{comp0}). Due to this symbol of combination  $K^{(1)}_{t_{-1}}A+K^{(1)}$ is bounded when $z\to\infty$. Thus there exists multiplication operator $V$ such, that 
\begin{equation}
K^{(1)}_{t_{-1}}A+K^{(1)}=V_{0}K.\label{dbar5}                                                                                                                                                                                                                                                                                                                                                                                                                                                                                                                                                                                                                                                       \end{equation}
Now, taking ``normal ordering'' (\ref{norm1}) into account, we decompose (\ref{dbar5}) with respect to $z$. Non-negative orders of these decomposition supply us with evolutions of the dressing operator. In particular we derive:
\begin{align}
&K^{(1)}_{t_{-1}}A+K^{(1)}_{} =(I+u^{(1)}_{t_{-1}})K,\label{t-1}\\
&K^{(1)}_{x_{-1}}A+iK^{(1)}_{}\sigma =(i\sigma+u^{(1)}_{x_{-1}})K,
\label{x-1}\\
&K^{(1)}A=KA+K_{t_1}+(u^{(1)}-u)K,\label{nn1}
\end{align}
where the last two equations are derived in analogy to (\ref{t-1}) and we took (\ref{norm1}) into account in the last line. 

\section{Jost solutions and compatibility conditions.}

The derivation of the compatibility conditions can be simplified by introducing the Jost solution instead of the dressing operator:
\begin{align}
\varphi(t_1,t_{-1},x_{-1},n,z)&=K(t_1,t_{-1},x_{-1},n,z)
{z}^n\exp\biggl(zt_1+
\dfrac{t_{-1}}{z}+i\sigma\dfrac{x_{-1}}{z}\biggr).\label{Jost}
\end{align}
This function has no meaning of tempered distributions, but enables to remove explicit dependence on $z$ in equations (\ref{t-1})--(\ref{nn1}):  
\begin{align}
&\varphi_{t_{-1}}^{(1)}=(I+u_{t_{-1}}^{(1)})\varphi,\label{b20}\\ 
&\varphi_{x_{-1}}^{(1)}=(i\sigma+u_{x_{-1}}^{(1)})\varphi.\label{b21}\\
&\varphi_{}^{(1)}=\varphi_{t_1}+(u_{}^{(1)}-u)\varphi,\label{b22}\end{align}
We see that (\ref{b22}) excludes shift of the Jost solution by  $n$ from the l.h.s.'s of (\ref{b20}) and (\ref{b21}):
\begin{align}
&\varphi_{t_1t_{-1}}+(u^{(1)}-u)\varphi_{t_{-1}}-(I+u_{t_{-1}})\varphi=0,\label{b23}\\ 
&\varphi_{t_1x_{-1}}+(u^{(1)}-u)\varphi_{x_{-1}}-(i\sigma+u_{x_{-1}})\varphi=0,\label{b24} 
\end{align}
that gives new dependent variable $u^{(1)}-u$. But we cannot resolve explicitly problem of compatibility of (\ref{b23}) and (\ref{b24}), because we cannot single out $\varphi_{t_{-1}}$ from (\ref{b23}), or  $\varphi_{x_{-1}}$ from (\ref{b24}) in order to check equality $\varphi_{t_{-1}x_{-1}}=\varphi_{x_{-1}t_{-1}}$. On the other hand, the existence of such a condition follows directly from the equalities (\ref{Kt-1}) and (\ref{Kx-1}), under condition of the unique solvability of the inverse problem. Indeed, differentiating the first of these equalities by $x_1$, we obtain by virtue of the second equality
\begin{align*}
&\overline\partial K_{t_{-1}x_{-1}}=K_{t_{-1}x_{-1}}B+iK_{t_{-1}}[\sigma A^{-1},B]+K_{x_{-1}}[A^{-1},B]+K[A^{-1},[\sigma A^{-1},B]],\\ 
&\overline\partial K_{x_{-1}t_{-1}}=K_{x_{-1}t_{-1}}B+K_{x_{-1}}[A^{-1},B]+iK_{t_{-1}}[\sigma A^{-1},B]+K[\sigma A^{-1},[A^{-1},B]],
\end{align*}
where the second equality is written out by analogy. Taking into account that the second and third terms in the right-hand sides overlap, and the difference of the last terms vanishes by the Jacobi identity, we get equality
\begin{equation*}
\overline\partial\bigl(K_{t_{-1}x_{-1}}-K_{x_{-1}t_{-1}}\bigr)=\bigl(K_{t_{-1}x_{-1}}-K_{x_{-1}t_{-1}})B, 
\end{equation*}
which, by virtue of (\ref{dbar_2}), proves the compatibility of these two evolutions.

On the other hand we can check directly compatibility of (\ref{b22}) and (\ref{b23}), i.e., equality 
$\bigl(\varphi_{t_{-1}}^{(1)}\bigr)^{(1)}=\bigl(\varphi^{(1)}\bigr)_{t_{-1}}^{(1)}$, that gives
\begin{align}
&u_{t_1t_{-1}}=(I+u_{t_{-1}})(u-u^{(-1)})-(u^{(1)}-u)(I+u_{t_{-1}}),
\label{ut}\\
\intertext{and compatibility of (\ref{b22}) and (\ref{b24}), i.e.,
$\bigl(\varphi_{x_{-1}}^{(1)}\bigr)^{(1)}=\bigl(\varphi^{(1)}\bigr)_{x_{-1}}^{(1)}$, that gives}
&u_{t_1x_{-1}}=(i\sigma+u_{x_{-1}})(u-u^{(-1)})-(u^{(1)}-u)(i\sigma+u_{x_{-1}}).\label{ux}
\end{align}
Both these equations present different versions of the Toda chain, but both of them involve only one evolution with the negative number, either $t_{-1}$, or $x_{-1}$. Neither $u^{(1)}$ nor $u^{(-1)}$ can be eliminated  from the system. 

We also can check the compatibility of (\ref{b20}) and (\ref{b21}), i.e., equality 
$(\varphi_{t_{-1}}^{(1)})^{(1)}_{x_{-1}}=(\varphi^{(1)}_{x_{-1}})^{(1)}_{t_{-1}}$. This gives
\begin{equation}
(I+u^{(1)}_{t_{-1}})(i\sigma+u_{x_{-1}})=(i\sigma+u^{(1)}_{x_{-1}})
(I+u_{t_{-1}}).\label{tx-1}
\end{equation}
Notice, that linearized version of this equation sounds as
\begin{equation}
i\omega^{(1)}_{t_{-1}}\sigma-\omega^{(1)}_{x_{-1}}-
i\sigma\omega_{t_{-1}}+\omega_{x_{-1}}=0,\label{o1} 
\end{equation}
that coincides with (\ref{b102}). Next, linearized versions of (\ref{ut}) and (\ref{ux}) give
\begin{equation}
\omega_{t_1t_{-1}}-2\omega+\omega^{(1)}-\omega^{(-1)}=0,\quad \omega_{t_1x_{-1}}-i\sigma\omega-i\omega\sigma-i\sigma\omega^{(1)}+i\omega^{(-1)}\sigma=0,\label{o2} 
\end{equation}
that coincides with (\ref{b10}) and (\ref{b100}), respectively. Moreover, linearized versions of equations (\ref{o1}) and (\ref{o2}) allow us to exclude shifted value of $\omega$, that gives 
\begin{equation}
(\omega_{t_{-1}t_{-1}}+\omega_{x_{-1}x_{-1}})_{t_1}-4\omega_{t_{-1}}=0,\label{o3}                                                                                                                                                                                                                     
\end{equation}
i.e., (\ref{b5}). Thus we arrive at an unusual situation: relations (\ref{b23}) and (\ref{b24}) are compatible, but the corresponding compatibility condition cannot be written explicitly in closed form. We need the additional variable $n$ to describe system with three independent variables $t_{-1}$, $x_{-1}$, and $t_{1}$. On the other hand, the derivation of the integrable equation (\ref{tx-1}) with independent variables $t_{-1}$, $x_{-1}$, and $n$ was based on the leading time $t_1$ and this time does not participate in the Lax pair (\ref{b20}), (\ref{b21}). Therefore, we need to choose another evolution as the leading one: either a discrete variable $n$ in (\ref{b3}), or times with negative numbers in (\ref{b6}).

\section{Different choices of leading evolutions.}
\subsection{Leading time $t_{-1}$.}
We start with a set of pseudo-differential operators $F$ with $t_{-1}$ as the leading time. We denote symbols of these operators as $F(t_{-1},x_{-1},n,z)$, etc. Variable $z$ cancels out from the commutator of the leading evolution, so instead of (\ref{b3}) we have here:
\begin{equation}
A^{-1}F=F_{t_{-1}}+z^{-1}F,\qquad FA^{-1}=z^{-1}F,\label{t100} 
\end{equation}
where $F$ is an arbitrary pseudo-differential operator and where $z\in\mathbb{C}$. Correspondingly the composition law (\ref{comp0}) must be changed to 
\begin{align}
(F&G)(t_{-1},x_{-1},n,z)=\nonumber\\
&=\dfrac{1}{2\pi}\int dp\int dy\,F\biggl(t_{-1},x_{-1},n,\dfrac{z}{1+ipz}\biggr)e^{ip(t_{-1}-y)}G(y,x_{-1},n,z),\label{comp3}
\end{align}
for any elements $F,G\in\mathcal{A}$, if the integral is well defined in the sense of distributions. Symbol of the unity operator is $I$ and symbol of the operator $A$ is the same as in (\ref{AF0}). But due to (\ref{comp3}) ``normal ordering'' for any $F$ now sounds as:
\begin{align}
(A^{-n}F)(t_{-1},x_{-1},n,z)&=(z^{-1}+\partial_{t_{-1}})^{n}F(t_{-1},x_{-1},n,z),\nonumber\\
(FA^{-n})(t_{-1},x_{-1},n,z)&=z^{-n}F(t_{-1},x_{-1},n,z),\label{AF10}
\end{align}
where $A^{-n}$ denotes the $n$-th power of $A^{-1}$ in the sense of composition (\ref{comp3}). In particular for $n=1$ we have (\ref{t100}) and for any $F$:
\begin{equation}
F_{t_{-1}}=[A^{-1},F].\label{F-1} 
\end{equation}

Notice that action of operator $A$ is more involved now. It is easier to present it in terms of the Fourier transform (\ref{f1}) but with respect to $t_{-1}$:
\begin{equation*}
\widetilde{(AF)}(p,x_{-1},n,z)=\dfrac{z}{1+ipz}
\widetilde{F}(p,x_{-1},n,z).
\end{equation*}
Correspondingly, now evolutions with respect to $x_{-1}$ and discrete variable $n$ (see (\ref{b6}) and (\ref{b3})) give in analogy to (\ref{b22}) and (\ref{b23})
\begin{equation}
\widetilde{B_{x_{-1}}} =i\sigma\biggl(\dfrac{2}{z}+ip\biggr)\widetilde{B}, \qquad 
\widetilde{B^{(1)}}=\dfrac{1}{1+2ipz}\widetilde{B}, \label{b200}
\end{equation}
but for the symbol of the  Scattering data we again have (\ref{Bf2}).

Definition of the Inverse problem is the same as in (\ref{dbar_1}) and (\ref{dbar_2}), where now composition in the r.h.s.\ of (\ref{dbar_1}) is understood in the sense of (\ref{comp3}). Again, we assume asymptotic decomposition (\ref{dbar4}), but equality (\ref{AF1}) is absent here because time $t_1$ is not the leading evolution. Instead, due to (\ref{t100}) we have (\ref{F-1})
for any $F$. Relations (\ref{t-1}) and (\ref{x-1}) follow as above from (\ref{dbar_1}), (\ref{dbar_2}), and (\ref{dbar4}). So we define the Jost solution (cf.\ (\ref{Jost})) by means of the equality
\begin{equation}
\varphi(t_{-1},x_{-1},n,z)=K(t_{-1},x_{-1},n,z)
{z}^n\exp\dfrac{t_{-1}+i\sigma x_{-1}}{z}.\label{Jost2}
\end{equation}
In this way we rederive the Lax pair (\ref{b20}), (\ref{b21}) and nonlinear integrable equation (\ref{tx-1}).
Choice of the variable $x_{-1}$ as the leading time leads to the similar construction. Thus we directly constructed equation with two times with negative numbers and one discrete variable. Let us consider another realization of this program.  

\subsection{The leading evolution by discrete variable.}
Now we consider shift of the discrete variable $n$ as the leading evolution. So we impose condition that (cf.\ (\ref{b6})) for an arbitrary operator $F$
\begin{equation}
F^{(1)}=AFA^{-1}.\label{t52} 
\end{equation}
Thus symbols of pseudo-differential operators depends on the same set of independent variables $t_{-1},x_{-1},n,z$, but definition of the composition law must be changed with respect to (\ref{comp0}) and (\ref{comp3}):
\begin{align}
(FG)&(t_{-1},x_{-1},n_1,z)=\nonumber\\
=&\oint_{|\zeta|=1}\dfrac{d\zeta}{2\pi 
i\zeta} F(t_{-1},x_{-1},n,z\zeta)\sum_{m\in\mathbb{Z}}\zeta^{n-m}G(t_{-1},x_{-1},m,z),\label{comp4}
\end{align}
where $\zeta\in\mathbb{C}$. In \cite{akp2008b} we have shown that this definition appears as result of the standard product of infinite matrices $F$ and $G$ in terms of their symbols. Operators with $\zeta$-independent symbols correspond to diagonal matrices. Symbols of the identity operator and operator $A$ are the same as in (\ref{Az0}). 

We use here the standard notations
\begin{equation}
\oint_{|\zeta|=1}\dfrac{d\zeta\,\zeta^{n}}{2\pi i\zeta}=\delta_{n,0},\qquad \delta _{c}^{}(\zeta)=\sum_{n=-\infty 
}^{\infty }\zeta^{n},\label{f36}
\end{equation}
where the latter one gives the delta-function on the unity contour, i.e.,
\begin{equation}
\oint\limits_{|\zeta|=1}\dfrac{d\zeta}{2\pi i\zeta}f(\zeta)\delta _{c}^{}(\zeta)=f(1)\label{36:1} 
\end{equation} 
for a test-function $f(\zeta)$. 

Under these definitions it is easy to check that for any operator $F$ composition law (\ref{comp4}) gives
\begin{align}
&(AF)(t_{-1},x_{-1},n,z)=zF(t_{-1},x_{-1},n+1,z)=zF^{(1)}(t_{-1},x_{-1},n,z),\label{prod30}\\
&(FA^{-1})(t_{-1},x_{-1},n,z)=\dfrac{1}{z}F(t_{-1},x_{-1},n,z),\label{prod31}\\
&(AFA^{-1})(t_{-1},x_{-1},n,z)=F(t_{-1},x_{-1},n+1,z),\text{ i.e., }AFA^{-1}=F^{(1)},\label{prod32}
\end{align}
that proves (\ref{t52}). As above we impose condition that symbols of these ``pseudo-matrix'' operators are tempered distributions with respect to their variables. 

The Dressing operator $K$ is defined by means of the Inverse problem (\ref{dbar_1}), (\ref{dbar_2}), where now composition in the r.h.s.\ of (\ref{dbar_1}) is understood in the sense of (\ref{comp4}), and we assume existence of the decomposition (\ref{dbar4}). Shift of the Dressing operator $K$ follow now from (\ref{t52}), 
\begin{equation}
K^{(1)}A=AK,\label{t-11}                                                                                                    
\end{equation}
so that the current version of the ``normal ordering'' takes the form $AK=K^{(1)}A$ instead of (\ref{norm1}) and (\ref{t100}). We see that relation (\ref{t-11}) contradicts (\ref{nn1}) due to the difference in the laws of composition (\ref{comp0}), (\ref{comp3}), and (\ref{comp4}). On the other hand, switching on the time $t_1$ by equality (\ref{Kt}), we get
\begin{equation*}
\overline\partial{K}_{t_1}=K_{t_1}B+KAB-KBA. 
\end{equation*}
The Inverse problem gives that there exists operator $Y$ with $z$-independent symbol, such that $K_{t_1}+KA=AK+YK$. Next, (\ref{dbar4}) show that $Y=u-u^{(1)}$, i.e., 
\begin{equation*}
K_{t_1}+KA=AK-(u^{(1)}-u)K. 
\end{equation*}
Here normal ordering is given by (\ref{t-11}), so that
\begin{equation}
K_{t_1}+KA=K^{(1)}A-(u^{(1)}-u)K,\label{2t}
\end{equation}
that coincides with (\ref{nn1}), after substitution $AK=K_{t_1}+KA$ there. 

\section{Conclusion.}
We have constructed two integrable systems, each of which has two independent variables with negative time numbers. In this case, the system with independent variables $t_1$, $t_{-1}$, $x_{-1}$ is defined only implicitly. More precisely, the compatibility conditions for the system (\ref{b23}) and (\ref{b24}), where $u^{(1)}-u$ denotes the new dependent variable, cannot be written out explicitly. Despite this, the evolution of the Scattering Data of this system is written out explicitly in (\ref{b5}), which give evidence that the system is integrable. However, both the solution of the Direct Scattering Problem and the construction of a complete set of integrals of motion require further consideration.

The second integrable system is explicitly defined by the Lax pair (\ref{b20}), (\ref{b21}) as a condition for its compatibility in (\ref{tx-1}). Two times with negative numbers, $t_{-1}$ and $x_{-1}$, and a discrete variable $n$ are used as independent variables here. A specific property of this system is the absence of 
$(1+1)$-dimensional integrable reductions. It is easy to see that they all boil down to equations that can be solved explicitly.

The author thanks S.\ Konstantinou-Rizos for fruitful remarks.

\section*{Funding.}
This work was supported by the Russian Science Foundation under grant no. 25-11-00081, 
https://rscf.ru/en/project/25-11-00081/.

\end{document}